\documentclass[pdflatex, sn-basic]{sn-jnl}

\usepackage{cite}

\usepackage[utf8]{inputenc}
\usepackage[T1]{fontenc}
\usepackage{graphicx}
\usepackage{booktabs}
\usepackage{subcaption}
\usepackage{url}
\usepackage{array}
\newcolumntype{T}{>{\tiny}p{5cm}}

\newcommand{\name}[1]{\textsc{#1}}
\newcommand{\estee}{\name{Estee}}

\newcommand{\tasks}{\ensuremath{\mathcal{T}}}
\newcommand{\outputs}{\ensuremath{\mathcal{O}}}
\newcommand{\arcs}{\ensuremath{\mathcal{A}}}

\newcommand{\ws}{\ensuremath{\mathcal{W}}}
\newcommand{\furl}[1]{\footnote{\url{#1}}}

\graphicspath{{./imgs/}}
\newcommand{\chartspath}{charts}
\newcommand{\shapespath}{shapes}

\jyear{2021}%

\setcitestyle{square,numbers,comma}

\begin{document}
\title{Analysis of Workflow Schedulers in Simulated Distributed Environments}

\author{\fnm{Jakub} \sur{Ber{\'a}nek}}\email{jakub.beranek@vsb.cz}
\equalcont{These authors contributed equally to this work.}
\author{\fnm{Stanislav} \sur{B{\"o}hm}}\email{stanislav.bohm@vsb.cz}
\equalcont{These authors contributed equally to this work.}
\author{\fnm{Vojt{\v e}ch} \sur{Cima}}\email{vojtech.cima@vsb.cz}
\equalcont{These authors contributed equally to this work.}

\affil{\orgname{IT4Innovations, VSB – Technical University of Ostrava},
\orgaddress{\city{Ostrava}, \country{Czech Republic}}}


\abstract{Task graphs provide a simple way to describe scientific workflows (sets of
tasks with dependencies) that can be executed on both HPC clusters and in the
cloud. An important aspect of executing such graphs is the used scheduling
algorithm. Many scheduling heuristics have been proposed in existing works;
nevertheless, they are often tested in oversimplified environments.
We provide an extensible simulation environment designed for prototyping and
benchmarking task schedulers, which contains implementations of various
scheduling algorithms and is open-sourced, in order to be fully reproducible.
We use this environment to perform a comprehensive analysis of workflow scheduling
algorithms with a focus on quantifying the effect of scheduling challenges that have
so far been mostly neglected, such as delays between scheduler invocations or partially
unknown task durations.
Our results indicate that network models used by many previous works might produce
results that are off by an order of magnitude in comparison to a more realistic model.
Additionally, we show that certain implementation details of scheduling algorithms which
are often neglected can have a large effect on the scheduler's performance, and they
should thus be described in great detail to enable proper evaluation.}


\maketitle

\section{Introduction}
\label{sec:introduction}

Representing a computation by a directed task graph is a
common programming model for defining programs for a distributed system or a
parallel computer. The main advantage of such a program description is the
possibility to capture parallelizable behavior of an application while allowing
to abstract the computation from specific architectures and computational
resources. Task graphs are becoming one of the most popular ways of executing
complex workflows on distributed systems (both cloud and high-performance clusters)
and it is an active research idea to design both new task execution frameworks
~\citep{dask,ray,legate,pygion,parsl} and scheduling algorithms~\citep{tarhan,onlinescheduling,wang2018list,schedwang}.

Task graphs are used with various levels of task
granularity. Fine-grained task graphs occur in the context of task-based
programming models where tasks are usually short running fragments of code
within a single program~\citep{Taxonomy2018,dagum1998openmp}. In
contrast, coarse-grained task graphs are used to represent complex workflows composed of
a set of potentially long-running programs~\citep{SciLuigi,cima2018hyperloom,cwl}. Although our
benchmarks primarily focus on the latter category, the results are generalizable to a wider
spectrum of task graph scheduling problems.

To execute a task graph as quickly as possible, it is crucial to produce a quality
schedule that will distribute the computation amongst multiple nodes to achieve
as much parallelization as possible, while also minimizing data transfers over the network.
Yet, finding the optimal schedule for a task graph is NP-hard even for very
restricted formulations (without transfer costs and resource
management)~\citep{Ullman1975}. Plenty of heuristics have been proposed to
tackle this problem, ranging from list-based scheduling to genetic algorithms.
Many surveys and comparisons of scheduling algorithms were published
in~\citep{hlfet1974,kwok1998benchmarking,hagras2003static,wang2018list}.


The primary objective of this paper is to analyze the behavior of various
scheduling heuristics and present the results in a verifiable and reproducible
form.
Most scheduler surveys assume an environment with an oversimplified
communication and computation model. Some works use more complex communication
models that attempt to simulate more realistic network
behavior~\citep{jarry2000dagsim,Macey1998,sinenncontention2005,tang2009communication}.
However, none of them deal with two important properties that inevitably arise
during the actual execution of real world task graphs, namely that the
scheduling itself takes time and that the duration of the individual tasks may
not be known in advance to the scheduler.
Also, to our best knowledge, no previous survey provides source codes of the
implemented schedulers. One of the findings of our analysis is that various
scheduler implementation details can have a large effect on the performance
of the scheduling algorithm. Surveys that do not provide detailed scheduler
source codes will thus be difficult to reproduce and verify.

It is not a goal of this paper to introduce new scheduling heuristics, rather
it should provide guidance on which scheduler implementation details should be
published and which benchmark properties should not be omitted in order to obtain
reproducible results.

This work has the following goals to improve the current situation:
\begin{itemize}
    \item Benchmark various scheduling algorithms in a complex communication
    and computation environment and provide the results in an open and
    reproducible form. This includes the task graphs, all source codes for
    schedulers and the simulation environment and also all benchmark scripts.

    \item Evaluate various simulated properties (such as network model or
    knowledge of task durations) to find out which have the largest effects on
    the performance of the individual schedulers.

    \item Provide an extensible simulation environment that facilitates
    prototyping and evaluation of task graph schedulers and network models.
\end{itemize}

This paper is structured as follows:
We describe the problem of task graph scheduling in
Section~\ref{sec:problem-statement}.
Section~\ref{sec:sota} gives a brief overview of related works.
Section~\ref{sec:simulation-environment} describes the simulation environment and implemented
scheduling algorithms.
Section~\ref{sec:benchmarks} describes benchmark methodology and benchmarked task graphs.
Section~\ref{sec:evaluation} contains benchmark results and discusses effects of
the simulated properties.
Lastly, we conclude in Section~\ref{sec:conclusions}.

%
%

\section{Problem statement}
\label{sec:problem-statement}

A \emph{task graph} is an acyclic graph where nodes represent tasks and output
data objects.
Formally, $\mathrm{TG} = (\tasks, \outputs, \arcs)$, where $\tasks$ is a set of
tasks, $\outputs$ is a set of data objects produced by tasks;
$\tasks \cap \outputs = \emptyset$.
$\arcs = (\tasks \times \outputs) \cup (\outputs \times \tasks)$ is a set of arcs between tasks
and objects. Let
$t \in \tasks, o \in \outputs$, then $(t, o) \in \arcs$ means that a task $t$ produces object $o$;
$(o, t) \in \arcs$ means that a data object $o$ is an input for task $t$. We
always assume that each object is produced by exactly one task
($\forall o \in \outputs: \lvert \arcs \cap (\tasks \times \{o\}) \rvert = 1$).
For a task $t$, we call the set $\{ o \in \outputs \mid (o, t) \in \arcs \}$
\emph{inputs of task $t$} and
$\{ o \in \outputs \mid (t, o) \in \arcs \}$ \emph{outputs of task $t$}.
We also assume that $(\tasks \cup \outputs, \arcs)$ forms a finite directed
acyclic graph.

Many works related to task graph scheduling assume that each task produces at
most one output; however, in practice having multiple outcomes from a single
task is a common requirement in workflows and is directly supported by some
frameworks (e.g. Luigi\furl{https://luigi.readthedocs.io/en/latest},
Rain\furl{https://github.com/substantic/rain}).
Multiple outputs per task can be simply modeled in systems supporting only one
output per task by introducing artificial tasks with zero execution times. Each
such task takes an output and decomposes it into pieces. However, as we do not
want to complicate scheduling by introducing dummy tasks that are actually not
necessary to schedule, our simulation environment directly supports tasks with
multiple outputs.

The task graph is executed on a set of \emph{workers}, processes/\-machines
that are able to execute tasks and produce their outputs. Let $\ws$ denote the
set of all workers.
When a task $t$ is finished on a worker $w$, all its output objects become
immediately available at worker $w$. The worker $w$ may send an object $o$ to
another worker $w'$ and make $o$ available on $w'$. A task $t$ can be executed
on worker $w$ only if all inputs of $t$ are available at worker $w$. We assume
that execution of each task is uninterruptible and non-replicable. We say that
a task $t$ is \emph{ready} if all its inputs are already computed; a task $t$
is \emph{enabled} on $w$ if $t$ is ready and all its inputs are available on
$w$.

The job of the scheduler is to assign tasks to workers, formally
to produce a map $S: \tasks \rightarrow \ws$. $Static$ schedulers produce
this map at the beginning of the computation and assign a worker to each task.
$Dynamic$ schedulers compose the map dynamically during the execution of the
task graph. The goal of the scheduler is to create $S$ such that it minimizes
the $makespan$ (the time it takes to finish all tasks in the graph).

A scheduler is allowed to change its decision and reschedule an already
scheduled task to a different worker. A task reschedule may fail if the task is
already running or if it has been already finished.

To align the simulation better with real-world task graph execution, we also
include the following properties:

\textbf{Multi-core workers}\quad Each worker may have multiple CPU cores; each
task may require a number of CPU cores. The total number of cores required by
simultaneously running tasks on a worker cannot exceed the total number of CPU
cores of that worker. This reflects the fact that currently most of commodity
and HPC processors have multiple CPU cores and software (represented by tasks)
can utilize them.

\textbf{Communication model}\quad
In many previous scheduler surveys and theoretical papers it is assumed that
the transfer time of a data object depends only on
the size of the object and not on the current network utilization
\citep{tang2010list,yao2013task,wang2018list,kwok1996dynamic}.
This is an unrealistic assumption about real computer networks, as the network
speed is affected by the number of concurrently running downloads. Moreover, it
is common that a real worker downloads more than one data object
simultaneously, which further affects the transfer durations because the
worker's bandwidth is shared.

We provide a more realistic network model that simulates full-duplex
communication between workers where the total upload and download bandwidth of
each worker is limited. The sharing of bandwidth between worker connections is
modeled by the \emph{max-min fairness model}~\citep{bertsekas_1992}. Max-min
fairness provides a bandwidth allocation for each worker. If we increase an
allocation of any participant, than we necessarily decrease the allocation of
some other participant with an equal or smaller allocation. When a download
starts or finishes, the data flow between workers is recomputed immediately,
thus we neglect the fact that it may take some time for the bandwidth to fully
saturate.

To compare this model with previous results, we also include the \emph{simple}
model in our simulation environment. It corresponds to the above mentioned
behavior used in several previous works. In our experiments we observe how the
makespan changes in response to the used network model.

\textbf{Worker inner scheduler}\quad
Since each worker has to keep track of its running tasks, manage resources, and
handle data object transfers, it becomes quite complex. In practice, the global
scheduler cannot micromanage each worker because this approach could not scale
to a larger number of workers. Therefore, we model a situation where each
worker has its own inner scheduler. We call it \emph{w-scheduler} and we
reserve the word ``scheduler'' for the global scheduler that assigns tasks to
workers.

The w-scheduler is not a subject of study in this work, hence we are going to
fix one particular worker scheduler and execute all experiments with it.
The implementation is inspired by the worker implementation used in
HyperLoom~\citep{cima2018hyperloom} and Rain. It is described in Appendix~A.


\textbf{Minimal scheduling delay}\quad
Dynamic schedulers create task assignments continuously, based on the current
situation. They could make a scheduling decision every time a task is finished;
however, in practice there is often an upper bound on the number of scheduler
invocations per second. It might be introduced artificially to reduce the
scheduling overhead or it might be caused by a software or hardware limitation
(e.g. messages with task changes cannot be received more often).
We introduce \emph{minimal scheduling delay} (MSD) that forces a minimal delay
between two scheduler invocations.

\textbf{Information modes}\quad
In most works, it is expected that the scheduler is aware of the duration of
all tasks and the sizes of all resulting data objects in advance. However, in
practice this information may not be available. In many cases, it may not be
clear for the author of the task graph how long will the tasks run or what will
be the size of the resulting objects (e.g. even for an experienced data
scientist, it may be hard to estimate how long will it take to train a machine
learning model on a particular dataset with particular hyperparameters). Even
if the task-graph author has precise knowledge of each task duration, it may be
tedious to manually annotate each task individually. Therefore, we consider the
following three modes of execution, which we call \emph{imodes}:

\begin{itemize}
    \item \emph{exact} -- scheduler has access to all task durations and object
    sizes for all elements in the task graph.
    \item \emph{user} -- for unfinished tasks, the scheduler has access only to
    a user-provided estimate of the task duration and its output sizes.
    \item \emph{mean} -- for unfinished tasks, the scheduler does not have any
    information about the duration or size of any graph element. However, the
    scheduler obtains the mean of the duration of all tasks and the mean of the
    size of all outputs.
\end{itemize}

Another possible scenario to consider could be a ``blind`` mode, where the
scheduler does not know any durations nor sizes in advance. However, in this
situation the schedulers would be very sensitive to an initial estimate of the
durations and sizes (namely the ratio between them, which influences decisions
whether to move data objects between workers). This estimate strongly
influences the early behavior of dynamic schedulers and it is completely vital
for static schedulers.
To avoid exploring various estimated values that would have to be chosen almost
arbitrarily, we propose to use the \emph{mean} mode instead of the blind mode.
We assume that if the scheduler knows nothing in advance, it could always
monitor the durations and sizes of finished tasks gradually and such monitored
values would converge to the mean. In practice, this would take some time, in
our environment the schedulers know about the mean in advance. Nevertheless, we
can often get a reasonable estimate of the mean durations based on previous
executions of similar workflows.

For the \emph{user} imode, we use values sampled from a random distribution
that is specific to a subset of tasks or objects within the task graph that
share similar properties (e.g. in MapReduce, all map operations use the same
distribution, all reduce operations use another distribution). Categorization
of tasks into these subsets was done manually. This simulates a user that is
able to categorize tasks and provide an estimate for each category.

In the experiments presented in this work, we aim to explore the behavior of
state of the art schedulers in a complex simulation environment that includes
all of the aspects described above.

Beside the comparison of individual schedulers, we also want to measure how
much does the used network model, information modes and minimal scheduling
delays affect the individual schedulers.
Many previous scheduler studies were performed in relatively simple
environments without these effects. We want to analyze whether there is a
significant difference between the performance of the standard heuristics when
they are benchmarked in more realistic conditions.

%
%
\section{Related Work}
\label{sec:sota}

Various workflow scheduling algorithms have been researched and implemented to
date (e.g. HLFET~\citep{hlfet1974}, SCFET~\citep{kwok1999static},
DLS~\citep{sih1993compile}, LAST~\citep{baxter1989last},
MCP~\citep{wu1990hypertool}, ETF~\citep{hwang1989scheduling}).
Number of publications overview and compare properties of these algorithms
\citep{kwok1998benchmarking,kwok1999static,wang2018list}.

Numerous surveys on distributed workflow environments and their schedulers have
been performed to categorize workflow environments based on their task
allocation strategies, load balancing, and multi-tenancy behaviour
\citep{hilman2018multiple,
surveytaskallocation, surveyemerging, taxonomy2016}. These are mostly focused
on cloud scenarios and scheduling algorithms are not their main focus. They
thus do not provide scheduler benchmarks.

Many works evaluate the algorithms using simplified environments with simple
communication models and without considering \emph{MSD} and \emph{imode}
effects.
In \citep{sinenncontention2005} a complex network model with various
network topologies was
considered, but it only reports results on two scheduling algorithms. The
\citep{tang2009communication} investigates the incorporation of contention
awareness into task scheduling. In \citep{Macey1998}, performance impact of
communication costs on static schedulers is studied.

All of these works use the assumption that task durations and data object sizes
are known in advance (i.e. in our terminology they use the \emph{exact} imode).
As far as we know, there was no systematic study of MSD or imodes in the
context of DAG scheduling.

Some of the popular distributed environment simulators such as
Simgrid~\citep{casanova2001simgrid} or CloudSim~\citep{calheiros2011cloudsim}
focus on deployment and provisioning infrastructures with low granularity of
resource requirements, but do not directly consider scheduling task workflows
with task dependencies.
This problem has been assessed by various tools built on top of these two
systems. DAGSim~\citep{jarry2000dagsim} only reports
experimental results without providing the actual implementation which makes it
difficult to extend the solution or reproduce the results.
SimDAG~\citep{zulianto2016hpc} does not consider task resource
requirements (e.g. number of cores) and also does not allow to define custom
network models. WorkflowSim~\citep{workflowsim},
ElasticSim~\citep{elasticsim}, CloudSim4DWf~\citep{cloudsim4dwf} and
Wrench~\citep{wrench} focus on simulating complex cloud scenarios, involving
datacenter costs, multi-tenancy, storage layers and other advanced factors.
Even though their simulation environments are very advanced, their scheduling
mostly operates on a different level of granularity, focusing on relatively
coarse-grained cloud or cluster jobs. Scheduling a large number of fine-grained
tasks is not their main focus and therefore it would be challenging to extend
their schedulers with support for MSD or imodes.

\section{Simulation environment}
\label{sec:simulation-environment}
This section describes the simulation environment that we have implemented to analyse
and compare various schedulers., benchmarked schedulers,
network models and task graph sets.

\subsection{Simulation}
We have implemented
\estee{}\footnote{\url{https://github.com/it4innovations/estee}},
a flexible open-source simulation environment that is designed for benchmarking
and experimenting with task schedulers. The implementation is very open-ended
and allows us to implement new schedulers, network models and workers easily.
However, it also comes ``battery-included`` and provides implementations for
all its components.

\subsection{Architecture}

\begin{figure}
	\centering
	\includegraphics[scale=0.25]{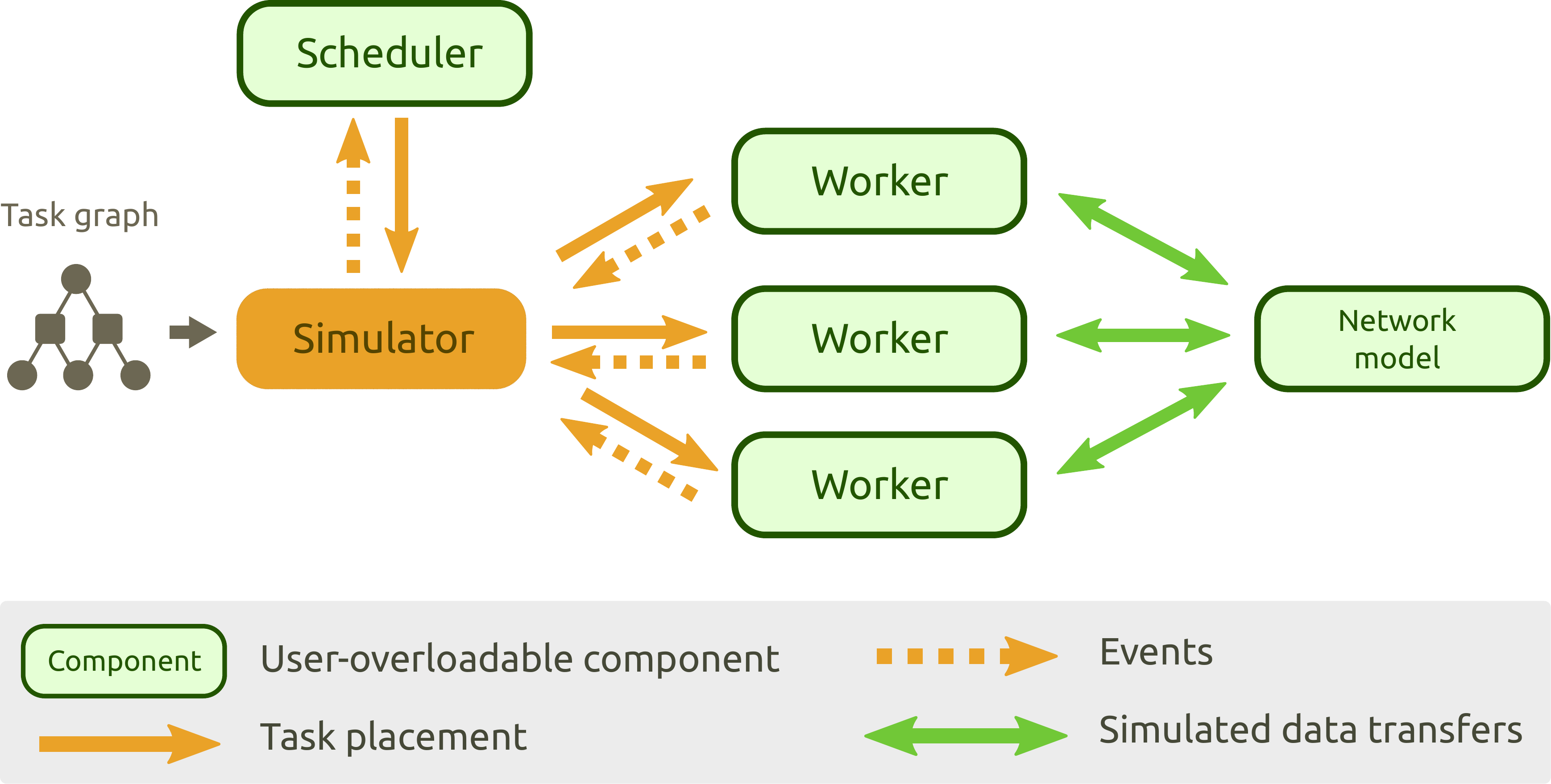}
	\caption{\estee{} architecture}
	\label{fig:arch}
\end{figure}

The architecture of \estee{} is depicted in Figure~\ref{fig:arch}. The central
component is the \emph{Simulator}, which controls the whole simulation and
communicates with the scheduler and workers.
The \emph{Scheduler} reads events about finished tasks and returns allocations
of tasks to workers.
A \emph{Worker} simulates the execution of assigned tasks and also the transfer
of task outputs between workers.
The communication between workers is handled by a network model that informs
them about download completion.


\estee{} is written in Python to provide a high-degree of flexibility that
facilitates rapid prototyping. \estee{} is an open-source project provided
under MIT license.

\subsection{Schedulers}
\label{sec:schedulers}
We have implemented a set of schedulers inspired by classic scheduling
heuristics. Originally these heuristics were mostly designed for environments
with only one core per worker and one output per task; therefore, we had to
slightly extend their implementation.

\noindent\textbf{blevel}\quad
Highest Level First with Estimated Times~\citep{hlfet1974} (HLFET) is a basic
list-based scheduling algorithm that prioritizes tasks based on their b-level.
B-level of a task is the length of the longest path from the task to any leaf
task (in our case the length of the path is computed using task durations,
without data object sizes). The tasks are scheduled in a decreasing order based
on their b-level.

\noindent\textbf{tlevel}\quad
Smallest Co-levels First with Estimated Times \citep{kwok1999static} is similar
to HLFET, with the exception that the value computed for each task (t-level) is
the length of the longest path from any source task to the given task. This
value corresponds to the earliest time that the task can start. The tasks are
scheduled in an increasing order based on their t-level.

\noindent\textbf{dls}\quad
Dynamic Level Scheduling \citep{sih1993compile} calculates a dynamic level for
each task-worker pair. It is equal to the static b-level lessened by the
earliest time that the task can start on a given worker (considering necessary
data transfers). In each scheduling step, the task-worker pair that maximizes
this value is selected.

\noindent\textbf{mcp}\quad
The Modified Critical Path \citep{wu1990hypertool} scheduler calculates the
ALAP (as-late-as-possible) time for each task. This corresponds to the latest
time the task can start without increasing the total schedule makespan. The
tasks are then ordered by this value in an ascending order and scheduled to the
worker that allows their earliest execution.

\noindent\textbf{etf}\quad
The ETF (Earliest Time First) scheduler \citep{hwang1989scheduling} selects the
task-worker pair that can start at the earliest time at each scheduling step.
Ties are broken by a higher static b-level.

\noindent\textbf{genetic}\quad
This scheduler implementation uses a genetic algorithm to schedule tasks to
workers. It uses the mutation and crossover operators described in ~\citep{omara2009genetic}.
Only valid schedules are considered, if no valid
schedule can be found within a reasonable amount of iterations, a random
schedule is generated instead.

\noindent\textbf{ws}\quad
Implementation of a simple work-stealing algorithm. The default policy is that
each ready task is always assigned to a worker where it can be started with
minimal transfer costs.
The scheduler monitors the load of workers and when a worker starts to starve
then a portion of tasks assigned to other workers is rescheduled to the
starving worker.

We have also implemented several naive schedulers to serve as a baseline for
scheduler comparisons.

\noindent\textbf{single}\quad
Scheduler that assigns all tasks to a single worker (it selects the worker with
the most cores). The resulting schedule never induces any data transfers
between workers.

\noindent\textbf{random}\quad
Static scheduler that schedules each task to a random worker.

All scheduler implementations use a random choice when an indistinguishable
decision in the algorithm occurs, e.g. when more tasks have the same b-level in
the case of \emph{blevel}.

We have implemented the list based schedulers (\emph{blevel}, \emph{tlevel},
\emph{dls}, \emph{mcp}, \emph{etf}) as closely as possible according to their
description from the works that introduced them. These heuristics often
schedule a task to a worker that allows the earliest start time of the task.
However, the scheduler algorithms do not prescribe in detail how exactly should
the scheduler find such worker, because the exact earliest start time often
cannot be determined in advance due to unpredictable network contention.
This implementation detail is crucial and should be included in the description
of new scheduling algorithms.

For our implementation, we used a simple estimation of the earliest start time
based on the currently running and already scheduled tasks of a worker and an
estimated transfer cost based on uncontended network bandwidth.

In addition, we have also created extended versions of the \emph{blevel},
\emph{tlevel} and \emph{mcp} schedulers to make them more compatible with the
additional properties that are present in our simulation environment (e.g.
multi-core workers, multi-core tasks, imodes). These modified versions use a
worker selection heuristic that we call ``greedy transfer`` and they contain
\emph{-gt} suffix in their name in the benchmark results. We have not applied
this heuristic to other schedulers, either because it could not be applied to
them without changing their behavior fundamentally or they already supported
the mentioned properties.

The ``greedy transfer`` heuristic assigns the selected task to a worker that
has a sufficient number of free cores on which the task may be executed and
that requires the minimal data transfer (sum over all sizes of data objects
that have to be transferred to that worker).
It also adds support for clusters where some machines have a different number
of cores than others. When a task $t$ that needs $c$ cores cannot be scheduled
because of an insufficient number of free cores, the list scheduling continues
by taking another task in the list instead of waiting for more free cores. This
task will only consider workers that have less than $c$ cores. This allows to
schedule more tasks while it does not modify the priority of tasks because $t$
cannot be scheduled on such workers anyway. Note that when all workers have the
same number of cores, the behavior is identical to ordinary list scheduling.

\section{Benchmark description}
\label{sec:benchmarks}
This section describes task graphs that we have used to compare the performance of
various scheduling algorithms and also cluster and scheduler configuration that
we have used in our benchmarks.

\subsection{Task graph datasets}

We use three task graph sets including simple elementary graphs as well as real
world inspired graphs to test the behavior of schedulers in various situations.
The first two sets are prepared by the authors, the third task graph set is
derived from a set commonly used in other works. All graphs are published
at~\citep{graphset2019}. \estee{} contains a task graph generator that can be used
to generate graphs from the following categories with various parametrizations.

\vspace{1mm}
\noindent\textbf{elementary}\quad contains trivial graph shapes, such as tasks
with no dependencies or simple fork-join graphs. This set should test how the
scheduler heuristics react to basic graph scenarios that frequently form parts
of larger workflows.

\vspace{1mm}
\noindent\textbf{irw}\quad is inspired by real world workflows, such as machine
learning cross-validation or map-reduce.

\vspace{1mm}
\noindent\textbf{pegasus}\quad is derived from graphs created by the Synthetic
Workflow Generators~\citep{pegasusgraphs}. The generated graphs correspond to
the
    \emph{montage}, \emph{cybershake}, \emph{epigenomics}, \emph{ligo} and
    \emph{sipht} workflows. We have extended the graphs with additional
    properties needed for testing imodes (notably expected task durations and
    data object sizes for the \emph{user} imode).

\vspace{1mm}
The properties of all used graphs are summarized in
Table~\ref{tab:graph_properties}.
Each task in all described task graphs requires at most four cores.

\begin{table*}
	\tiny
	\caption{Task graph properties}
	\centering
	\label{tab:graph_properties}
\begin{tabular}{l|lrrrr|T}
	\toprule
	Graph & D &   \#T &     \#O & TS &   LP & Description \\
	\midrule
	plain1n &   e &  380 &      0 &    0.00 &    1 &   Independent tasks;
	normally distributed durations (Fig.~\ref{fig:tg-plain}) \\
	plain1e &   e &  380 &      0 &    0.00 &    1 &    Independent tasks;
	exponentially distributed durations (Fig.~\ref{fig:tg-plain}) \\
	plain1cpus &   e &  380 &      0 &    0.00 &    1 &  Independent tasks with
	varying core requirements (Fig.~\ref{fig:tg-plain}) \\
	triplets &   e &  330 &    220 &   17.19 &    3 & Task triplets; middle
	task requires 4 cores (Fig.~\ref{fig:tg-triplets}) \\
	merge\_neighb. &   e &  214 &    107 &   10.36 &    2 & Merge of adjacent
	task pairs (Fig.~\ref{fig:tg-w}) \\
	merge\_triplets &   e &  148 &    111 &   10.77 &    2 & Merge of task
	triplets (Fig.~\ref{fig:tg-merge-triplets}) \\
	merge\_sm-big &   e &  240 &    160 &    7.74 &    2 &  Merge of two
	results (0.5 MiB and 100 MiB data objects) (Fig.~\ref{fig:tg-v}) \\
	fork1 &   e &  300 &    100 &    9.77 &    2 &      Tasks with a pair of
	consumers each consuming the same output (Fig.~\ref{fig:tg-fork})  \\
	fork2 &   e &  300 &    200 &   19.53 &    2 &      Tasks with a pair of
	consumers each consuming different output (Fig.~\ref{fig:tg-fork2}) \\
	bigmerge &   e &  321 &    320 &   31.25 &    2 &    Merge of a large
	number of tasks (variant of Fig.~\ref{fig:tg-merge}) \\
	duration\_stairs &   e &  380 &      0 &    0.00 &    1 &    Independent
	tasks; task durations range from 1 to 190 s (Fig.~\ref{fig:tg-plain}) \\
	size\_stairs &   e &  191 &    190 &   17.53 &    2 &  1 producer 190
	outputs / 190 consumers; sizes range from 1 to 190 MiB \\
	splitters &   e &  255 &    255 &   32.25 &    8 &  Binary tree of
	splitting tasks (Fig.~\ref{fig:tg-splitters}) \\
	conflux &   e &  255 &    255 &   31.88 &    8 &    Merging task pairs
	(inverse of \emph{splitters}) (Fig.~\ref{fig:tg-conflux}) \\
	grid &   e &  361 &    361 &   45.12 &   37 & Tasks organized in a 2D grid
	(i.e. \emph{splitters} followed by \emph{conflux}) (Fig.~\ref{fig:tg-grid})
	\\
	fern &   e &  401 &    401 &   11.11 &  201 &       Long task sequence with
	side tasks (Fig.~\ref{fig:tg-fern}) \\ \hline
	gridcat &   i &  401 &    401 &  115.71 &    4 & Merge of pairs of 300 MiB
	files  \\
	crossv &   i &   94 &     90 &    8.52 &    5 &  Cross validation \\
	crossvx &   i &  200 &    200 &   32.66 &    5 & Several instances of cross
	validation \\
	fastcrossv &   i &   94 &     90 &    8.52 &    5 & Same as \emph{crossv}
	but tasks are $50\times$ shorter \\
	mapreduce &   i &  321 &  25760 &  439.06 &    3 & Map-reduce pattern \\
	nestedcrossv &   i &  266 &    270 &   28.41 &    8 & Nested cross
	validation \\ \hline
	montage &   p &   77 &    150 &    0.21 &    6 &        Montage workflow
	from Pegasus \\
	cybershake &   p &  104 &    106 &    0.84 &    4 &        Cybershake
	workflow from Pegasus \\
	epigenomics &   p &  204 &    305 &    1.36 &    8 &        Epigenomics
	workflow from Pegasus \\
	ligo &   p &  186 &    186 &    0.11 &    6 &        Ligo workflow from
	Pegasus \\
	sipht &   p &   64 &    136 &    0.12 &    5 &        Sipht workflow from
	Pegasus \\
	\bottomrule
\end{tabular}\\
\vspace{2mm}
D = Dataset (e = elementary, i = irw, p = pegasus);
\#T = Number of tasks; \#O = Number of outputs;
TS = Sum of all output object sizes (GiB);
LP = longest oriented path in the graph
\end{table*}

\begin{figure}
	\centering
	\begin{subfigure}{.12\textwidth}
		\centering
		\includegraphics[width=.8\linewidth]{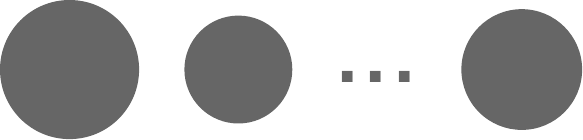}
		\caption{}
		\label{fig:tg-plain}
	\end{subfigure}%
	\begin{subfigure}{.12\textwidth}
		\centering
		\includegraphics[width=.8\linewidth]{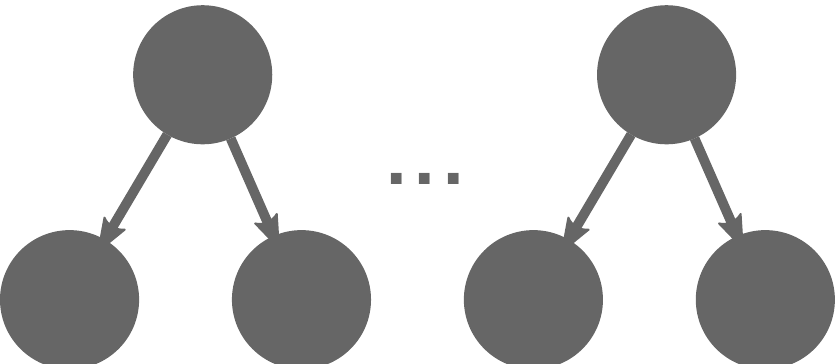}
		\caption{}
		\label{fig:tg-fork}
	\end{subfigure}
	\begin{subfigure}{.12\textwidth}
		\centering
		\includegraphics[width=.8\linewidth]{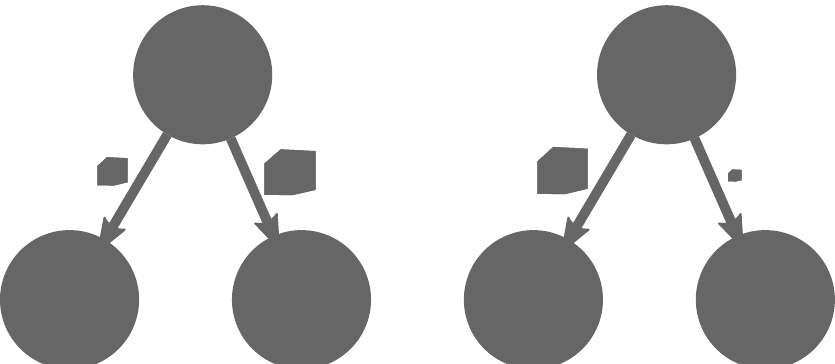}
		\caption{}
		\label{fig:tg-fork2}
	\end{subfigure}
	\begin{subfigure}{.12\textwidth}
	    \centering
	    \includegraphics[width=.8\linewidth]{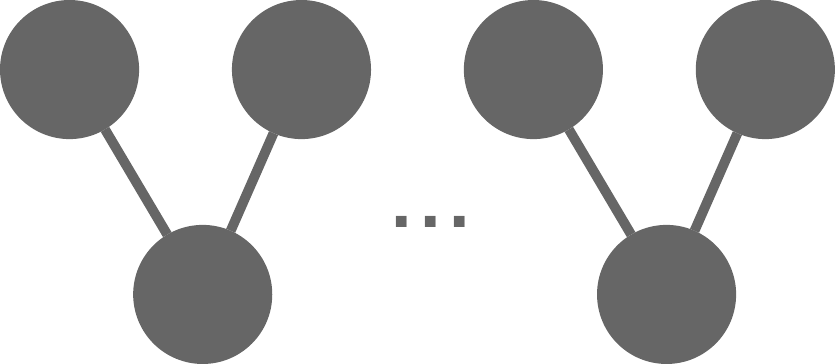}
	    \caption{}
	    \label{fig:tg-v}
    \end{subfigure}
\\
	\begin{subfigure}{.12\textwidth}
	    \centering
	    \includegraphics[width=.8\linewidth]{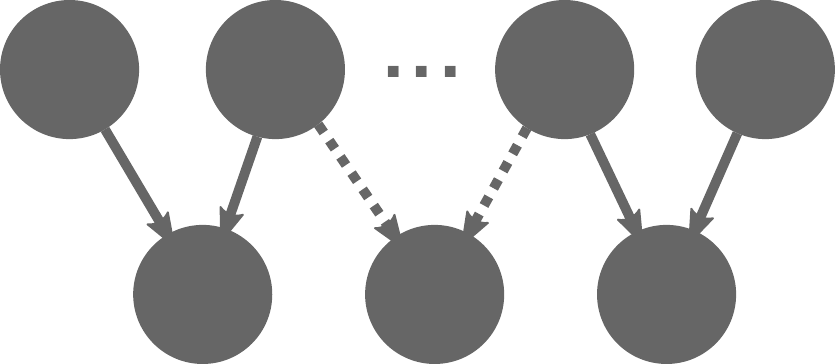}
	    \caption{}
	    \label{fig:tg-w}
    \end{subfigure}
	\begin{subfigure}{.12\textwidth}
	    \centering
	    \includegraphics[width=.8\linewidth]{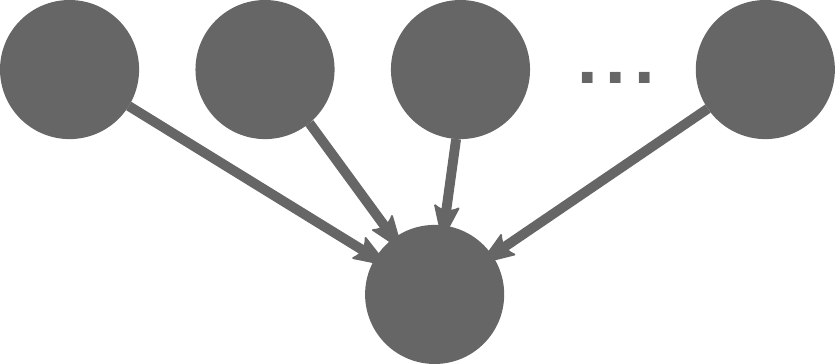}
	    \caption{}
	    \label{fig:tg-merge}
    \end{subfigure}
	\begin{subfigure}{.12\textwidth}
		\centering
		\includegraphics[width=.8\linewidth]{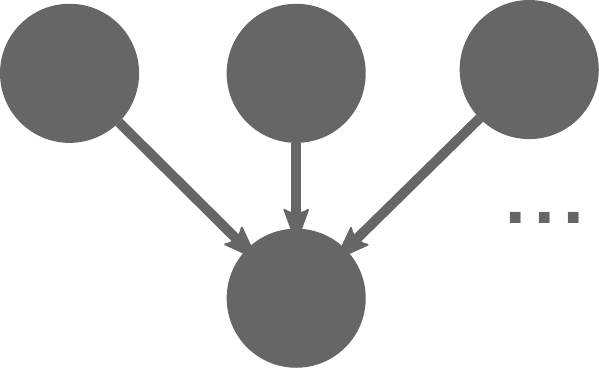}
		\caption{}
		\label{fig:tg-merge-triplets}
	\end{subfigure}
	\begin{subfigure}{.12\textwidth}
		\centering
		\includegraphics[width=.8\linewidth]{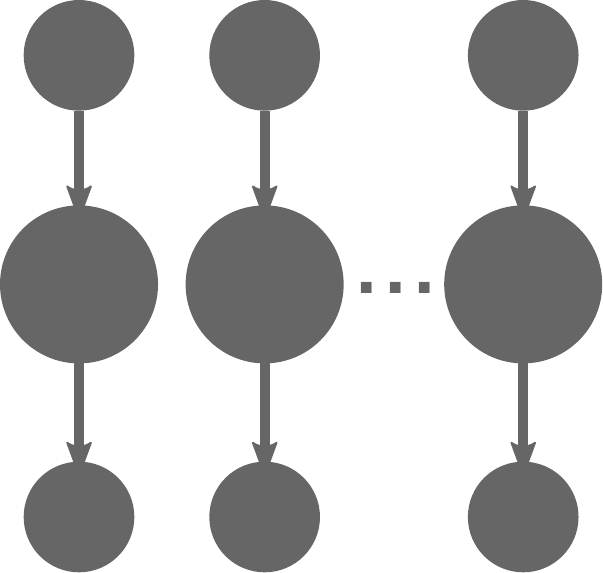}
		\caption{}
		\label{fig:tg-triplets}
	\end{subfigure}
\\
	\begin{subfigure}{.12\textwidth}
	    \centering
	    \includegraphics[width=.8\linewidth]{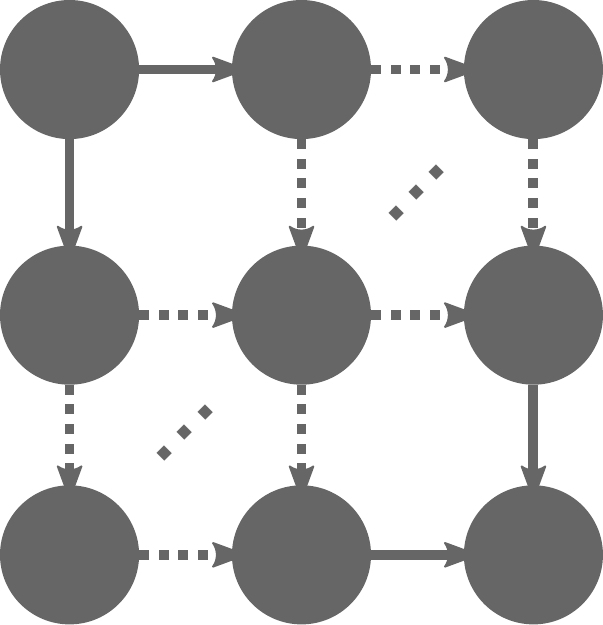}
	    \caption{}
	    \label{fig:tg-grid}
    \end{subfigure}
	\begin{subfigure}{.12\textwidth}
	    \centering
	    \includegraphics[width=.8\linewidth]{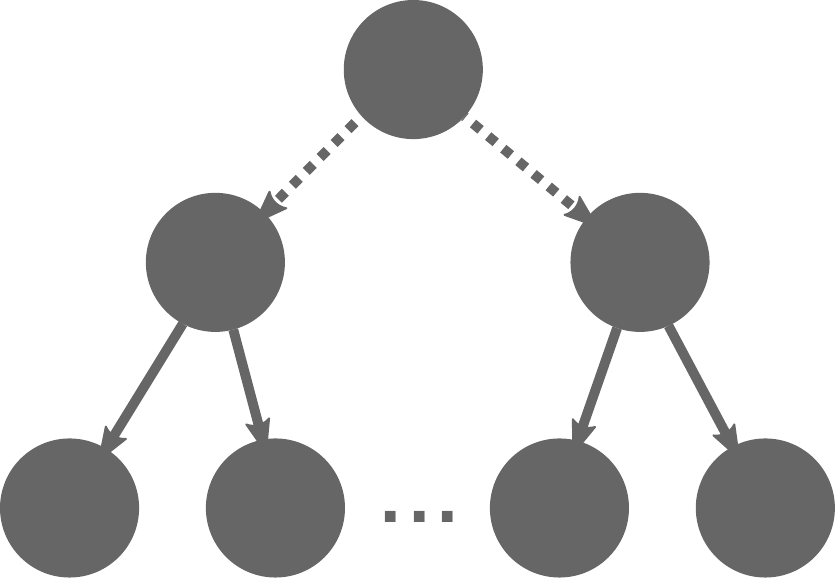}
	    \caption{}
	    \label{fig:tg-splitters}
    \end{subfigure}
	\begin{subfigure}{.12\textwidth}
		\centering
		\includegraphics[width=.8\linewidth]{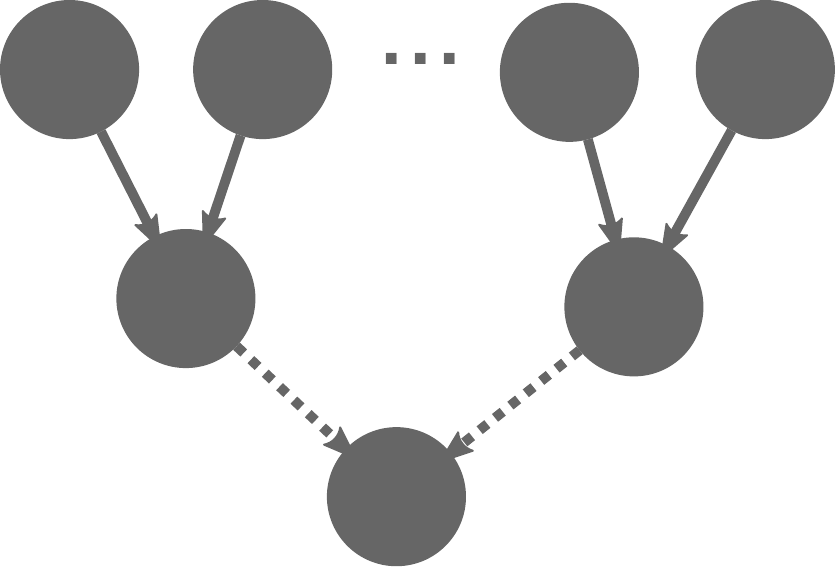}
		\caption{}
		\label{fig:tg-conflux}
	\end{subfigure}
	\begin{subfigure}{.12\textwidth}
		\centering
		\includegraphics[width=.8\linewidth]{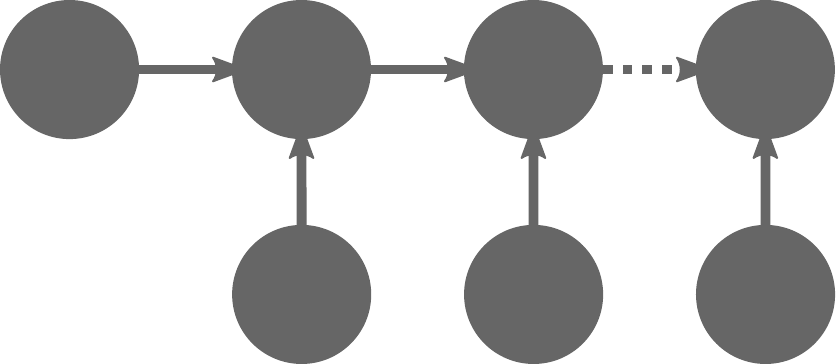}
		\caption{}
		\label{fig:tg-fern}
	\end{subfigure}

	\caption{Task graph shapes in the \emph{elementary} data set}
	\label{fig:tg-shapes}
\end{figure}

\subsection{Clusters}
We have used the following cluster configurations (where $w \times c$ means
that the cluster has $w$ workers and each worker has $c$ cores):  8$\times$4,
16$\times$4, 32$\times$4, 16$\times$8, 32$\times$16.

For simulating network connections, we use the \emph{max-min fairness} and
\emph{simple} network models with bandwidths ranging from 32 MiB/s to 8 GiB/s.
For experiments that do not focus on the network model (e.g. comparing imodes),
we only use the \emph{max-min} network model.

\subsection{Scheduler settings}
For evaluating the effect of MSD, we benchmark several MSD configurations.
As a baseline we use a configuration with no delay (MSD is zero), i.e. the
scheduling process is executed as soon as an event occurs.
Beside the base case we have also measured delays of 0.1, 0.4, 1.6, and 6.4
seconds. In all these non-zero cases, we have also added a 50 milliseconds
delay before sending the scheduler decision to workers to simulate the
scheduler computation delay. For experiments that do not focus on MSD, we
always use MSD of 0.1 seconds and 50 milliseconds delay.

For testing the effect of imodes, we benchmark schedulers with different
information modes (exact, user, and mean) as defined in
Section~\ref{sec:problem-statement}. For experiments that do not focus on
imodes, we always use the \emph{exact} imode.

%
%
\section{Evaluation}
\label{sec:evaluation}


%

This section discusses results obtained by running the described benchmarks in
our simulation environment. All obtained results are published
at~\citep{results2019} including the generated charts for all configurations.
Each particular configuration described in the
previous section was executed 20 times, except for the \emph{single} scheduler,
which was executed only once, since it is deterministic. Unless specified otherwise,
the experiments were performed with the default benchmark configuration
(\emph{max-min} netmodel, \emph{exact} imode and 0.1s MSD).

Our benchmarks have produced large amounts of results. Below we discuss several
noteworthy results, you can find more complete scheduler comparison results in the appendix.

\begin{figure*}
	\caption{Random scheduler performance}
	\label{f:random-scheduler}
	\centering
\includegraphics[width=\textwidth]{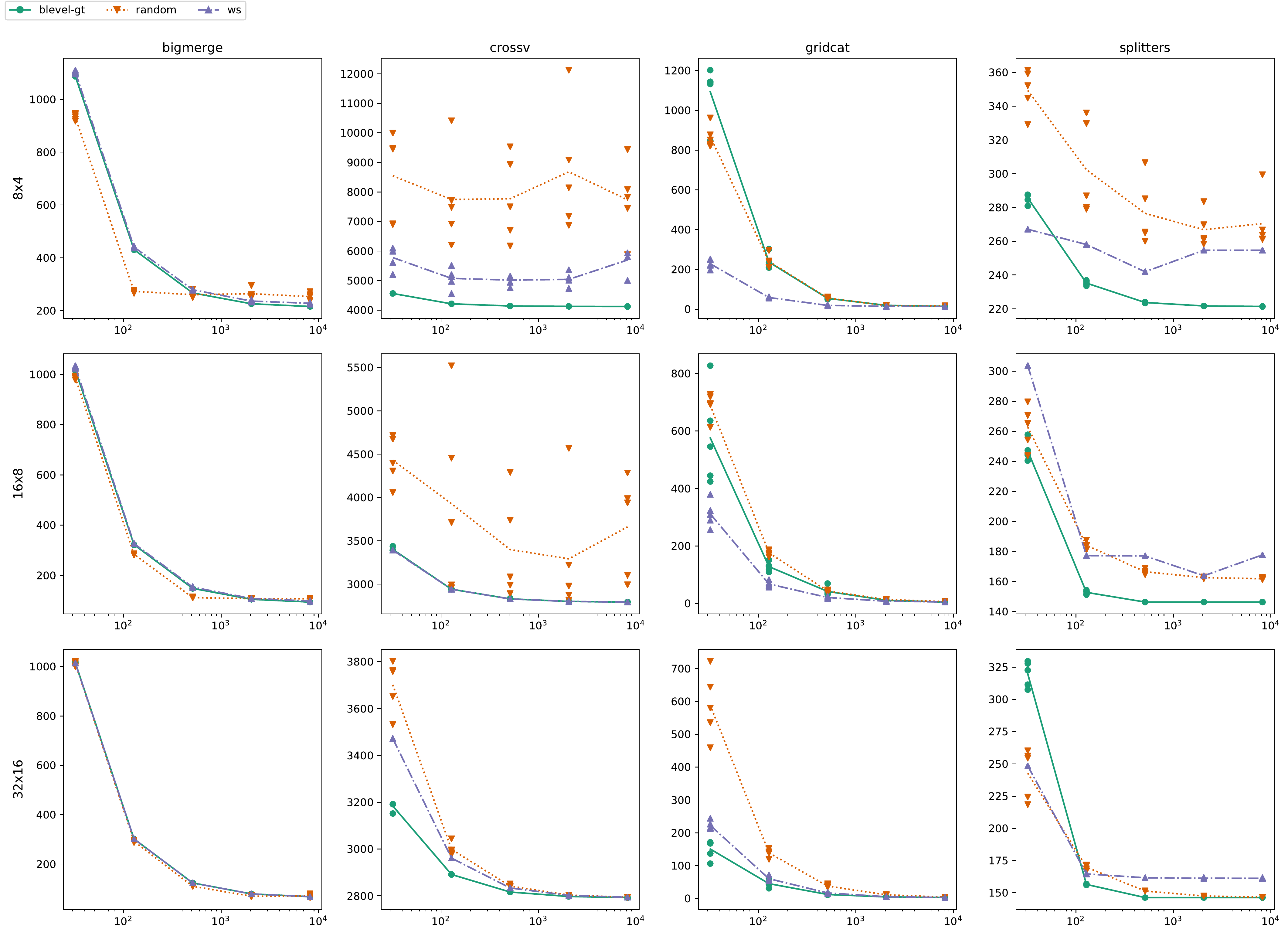}\\
{\small x axis: bandwidth [MiB/s]; y axis: makespan [s]}
\end{figure*}

\textbf{Random scheduler}\quad Figure~\ref{f:random-scheduler} shows simulated
makespan lengths produced by the \emph{random} scheduler and two other competitive
schedulers, \emph{blevel-gt} and the workstealing scheduler on selected graphs.
While the random scheduler produces quite long makespans in certain cases (for example
in the cross-validation graph), it is also surprisingly often quite competitive. Especially
as the number of workers and the bandwidth increases, it can get even with other schedulers
and sometimes even overcome them.

Similar results have also been observed in~\citep{rsds}. These results show that as the
computational cluster and network transfer speed gets larger, scheduling decisions can become less
important and other factors (like the runtime overhead of the task execution system) can start to
dominate.

\begin{figure*}
	\caption{Comparison of worker selection strategy}
	\label{f:gt-scheduler}
	\centering
\includegraphics[width=0.8\textwidth]{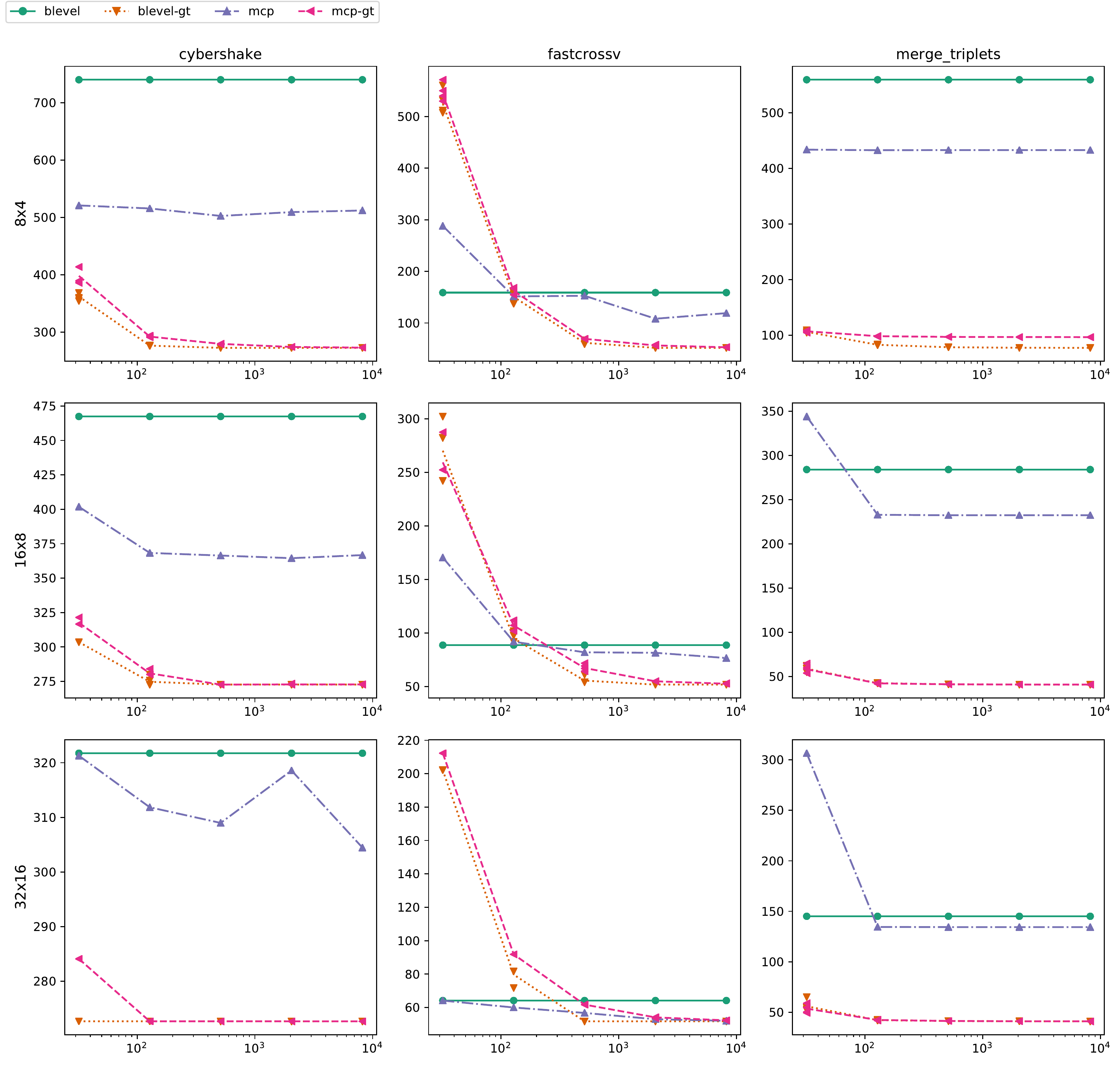}\\
{\small x axis: bandwidth [MiB/s]; y axis: makespan [s]}
\end{figure*}

\textbf{Worker selection strategy}\quad We have already explained in Section~\ref{sec:schedulers}
that published scheduler algorithms do not always specify the exact implementation of worker selection.
Yet as we can see in Figure~\ref{f:gt-scheduler}, this implementation detail is crucial. The worker
selection strategy (which is the only thing that differentiates the schedulers with and without
the \texttt{-gt} suffix) has a large effect on the produced schedule and thus the resulting
makespan. Furthermore, it is evident that schedulers that use the ``greedy transfer`` selection
strategy are highly correlated, which hints that in this case selecting the correct workers is more
important than scheduling the order in which tasks will be executed.

\begin{figure*}
	\caption{Total transfers on IRW dataset}
	\label{f:irw-transfer}
	\centering
\includegraphics[width=0.8\textwidth]{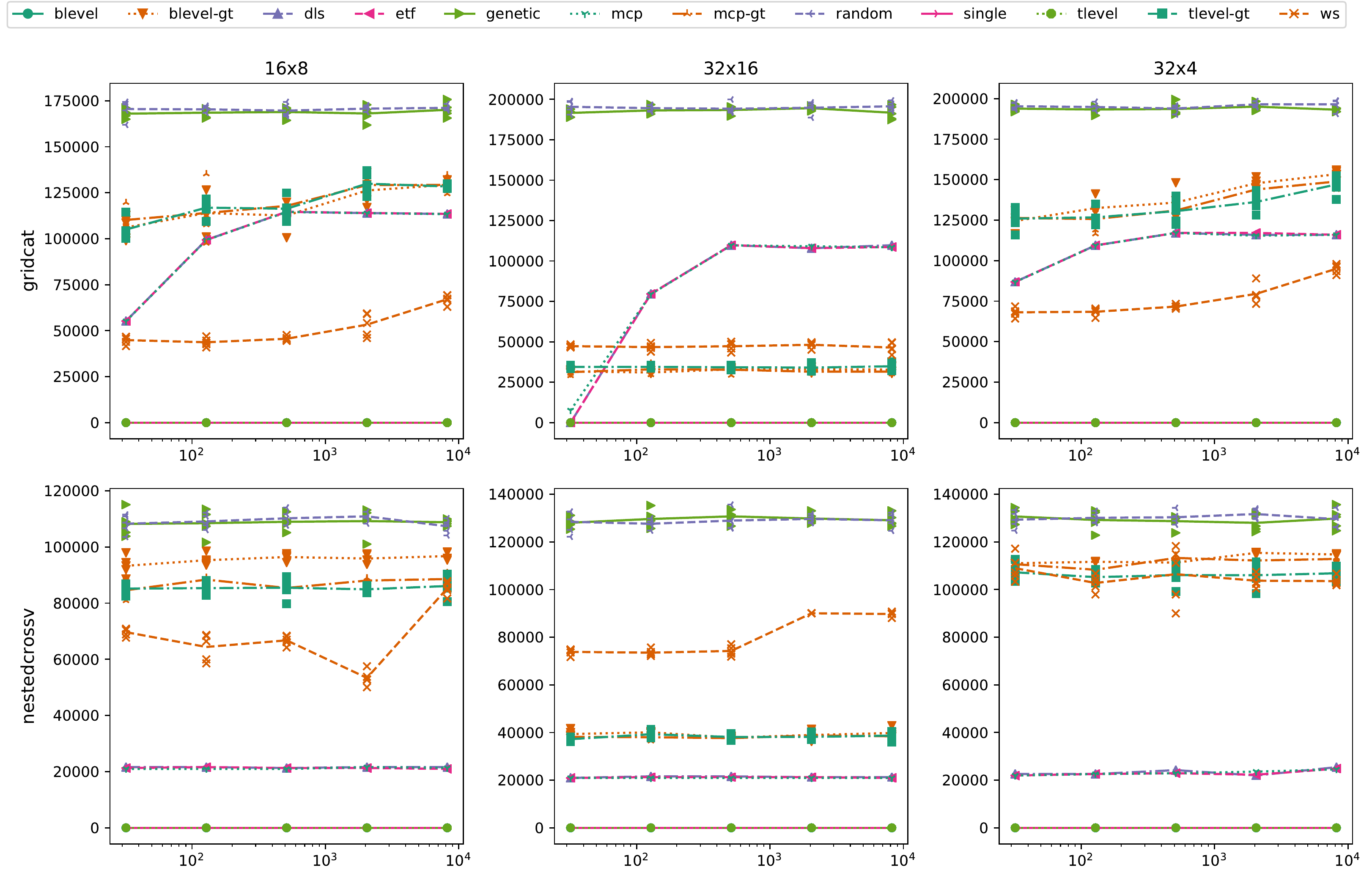}\\
{\small x axis: bandwidth [MiB/s]; left y axis: makespan [s], right y axis: sum of transferred data
	between all workers [MiB]}
\end{figure*}

\textbf{Network transfers}\quad Figure~\ref{f:irw-transfer} demonstrates that schedulers producing similar
makespans may in fact generate vastly different amounts of network traffic. For
example, for the \emph{nestedcrossv} graph using the \emph{32x16} cluster, the
work stealing scheduler transfers almost twice as much data than
\emph{blevel-gt}, yet it produces almost identical makespans.

%
%


\begin{figure*}
	\caption{Comparison of ``max-min'' and ``simple'' netmodel on IRW set;
	cluster 32x4}
	\label{f:irw-netmodel}
	\centering
\includegraphics[width=\textwidth]{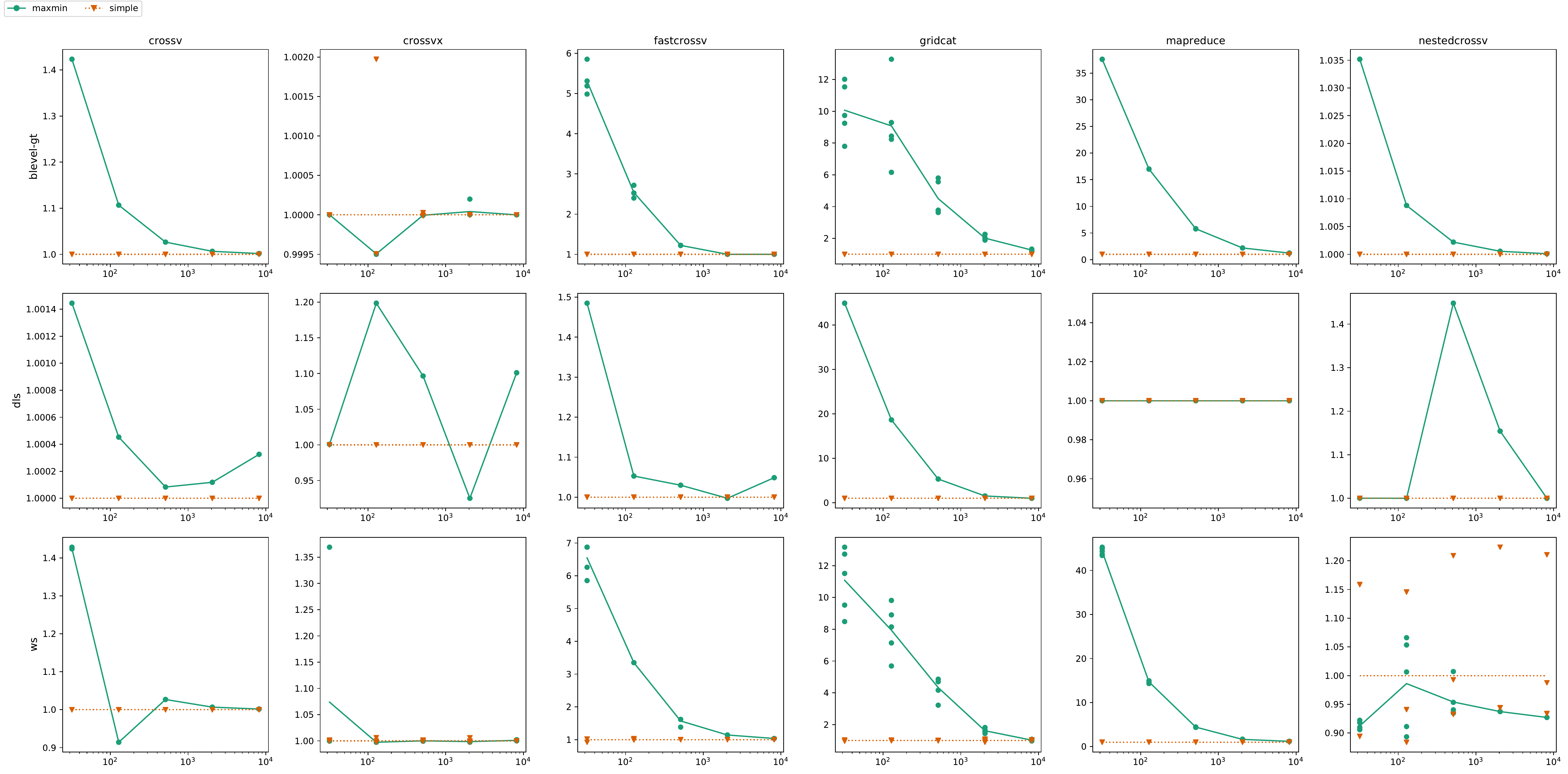}\\
{\small x axis: bandwidth [MiB/s]; y axis: makespan normalized to average
	of makespan of ``simple'' model}
\end{figure*}

\textbf{Network models}\quad
Figure~\ref{f:irw-netmodel} compares makespans between the \emph{simple} and
\emph{max-min} network models on the IRW
data set using the 32$\times$4 cluster for selected schedulers. The results are
normalized with respect to the \emph{simple}
model. It is clear that results obtained by using the \emph{simple} model often
under-approximate the resulting makespan
length. This is caused by the fact that network contention is not taken into
account, which causes their transfer duration
estimation to be overly optimistic. It is however interesting to note that in
some cases the \emph{simple} model
over-approximates the makespan. Since most of the schedulers use heuristics, a
faster network transfer does not necessarily
lead to a shorter makespan.

On the IRW dataset, the differences vary based on the particular scheduler and
task graph. Especially with slower bandwidths, the estimations produced by the
two models can be an order of magnitude apart. Note that even small disparities
are significant, since as shown in previous surveys~\citep{wang2018list} and in
our provided results, the differences in produced makespans between existing
scheduler heuristics are often very small and within a factor of two. As the
bandwidth gets faster, the difference between the two models decreases, since
network contention is lower and the \emph{max-min} model starts to behave
similarly to the \emph{simple} model.

For the Pegasus data set, the results of both models are much more aligned.
Differences on higher bandwidths are almost negligible. For slower bandwidths,
the differences between the models are within a factor of two. Results for the Pegasus data set
can be observed in Figure~\ref{f:pg-netmodel} in the appendix.

\begin{figure}
	\caption{Comparison of MSD on IRW subset; cluster 32x4}
	\label{f:irw-msd}
	\centering
\includegraphics[width=0.6\textwidth]{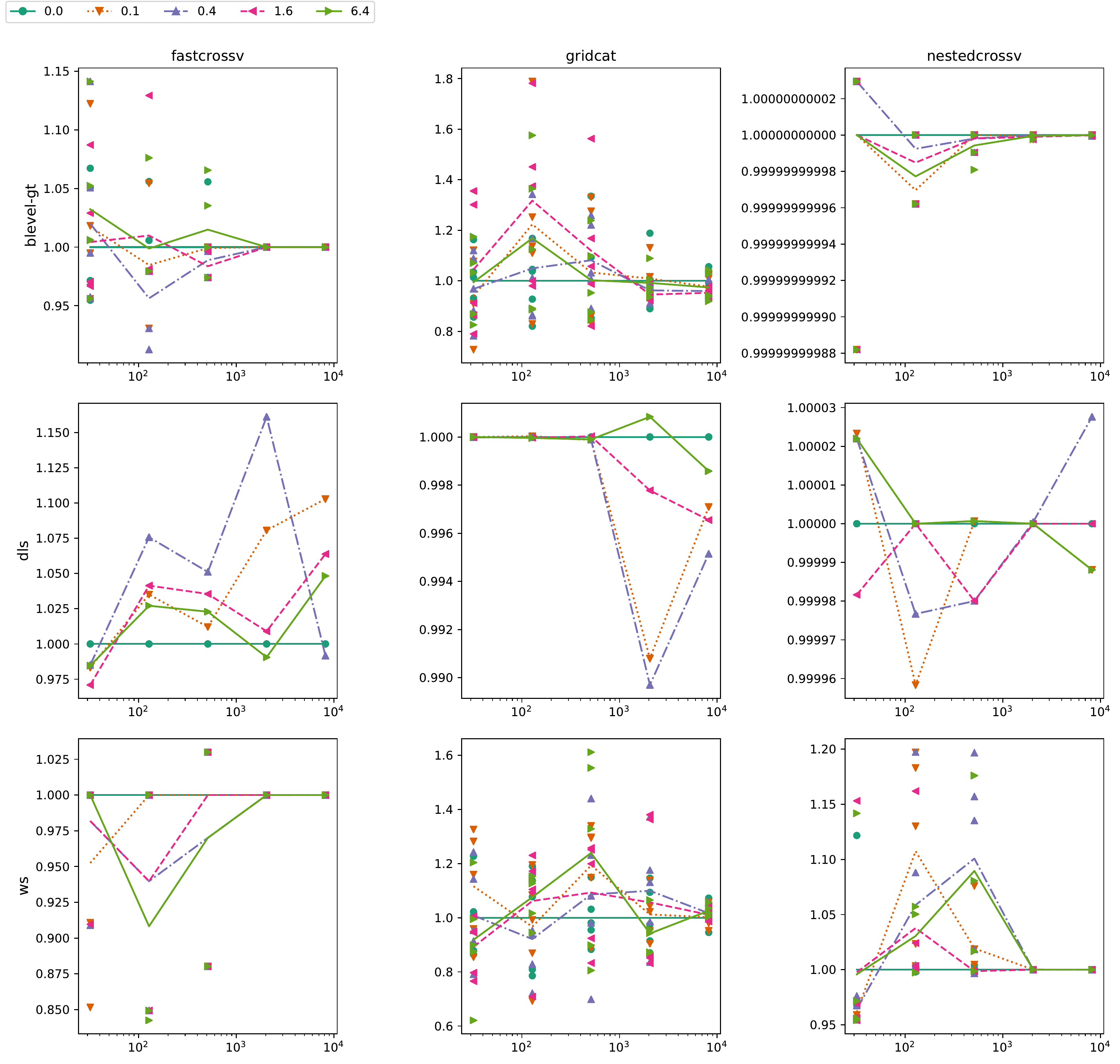}\\
{\small x axis: bandwidth [MiB/s]; y axis: makespan normalized to MSD 0.0
case}
\end{figure}

\textbf{MSD}\quad
Figure~\ref{f:irw-msd} shows the effect of MSD on the IRW data set using the
\emph{32x4} cluster for selected schedulers. The results are normalized with
respect to the case where MSD equals zero.


Our results show that the effect of MSD is relatively limited, especially when
compared to the effect of the simulated network model. There seems to be no
clear pattern as to whether decreasing MSD improves the makespan length
consistently or not. It is however interesting to note that increasing MSD can
actually improve the produced schedules (e.g. the \emph{ws} scheduler on the
\emph{fastcrossv} graph). Increasing the delay between individual scheduling
decisions introduces a ``batching`` effect. Even though the scheduler is
allowed to make decisions less often, it has access to more accumulated events
that happened in the meantime and it can thus potentially make a better
decision. Using an artificial MSD in a real scheduler implementation can thus
serve to improve the produced schedules, not just to reduce the scheduling
overhead.

\begin{figure*}
	\caption{Comparison of imodes on IRW set; cluster 32x4}
	\label{f:irw-imode}
	\centering
\includegraphics[width=\textwidth]{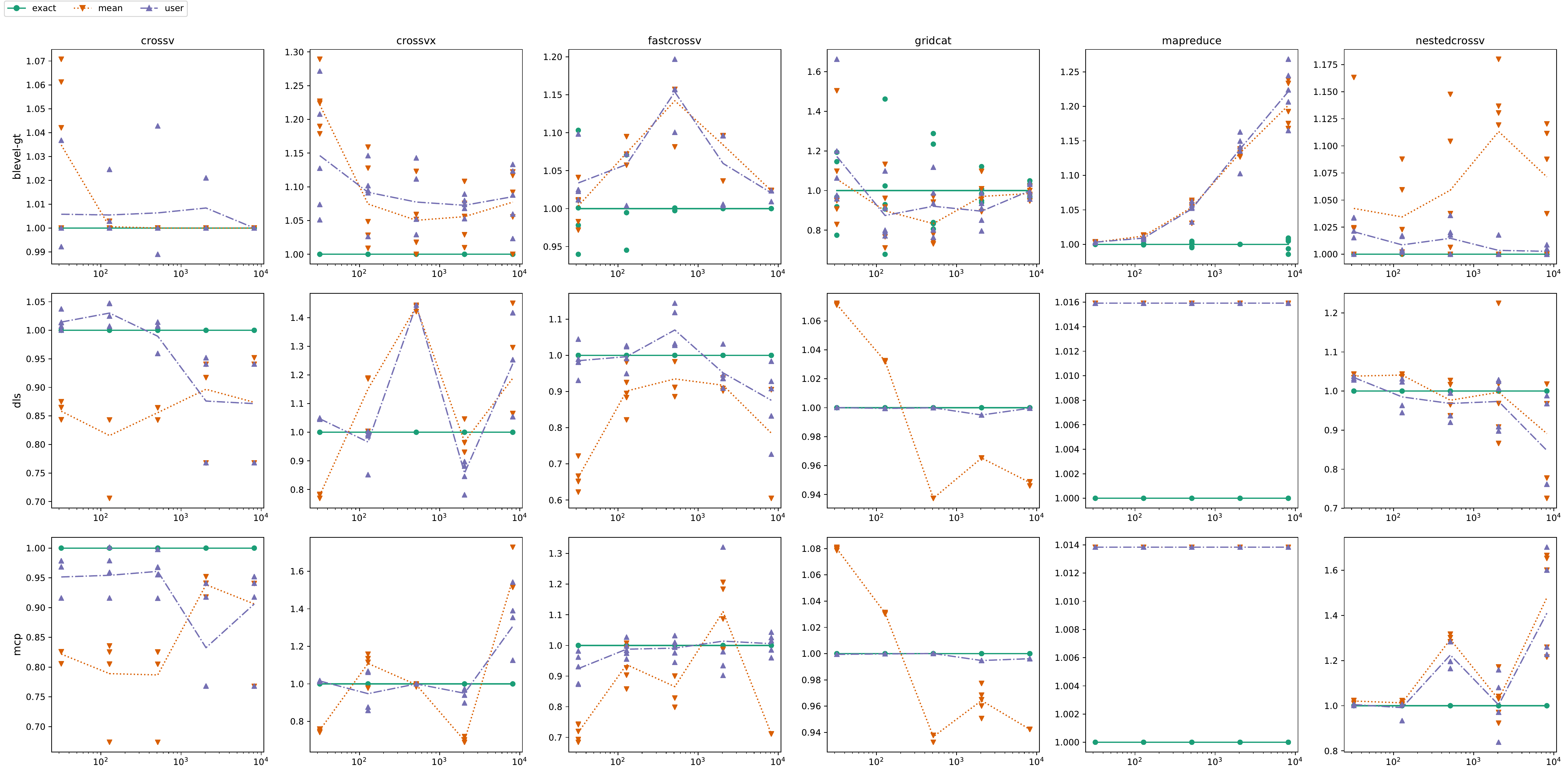}\\
{\small x axis: bandwidth [MiB/s]; y axis: makespan normalized to ``exact''
imode}
\end{figure*}

\begin{figure}
	\caption{Comparison of imodes on three elementary graphs; cluster 32x4}
	\label{f:elementary-imode}
	\centering
\includegraphics[width=0.6\textwidth]{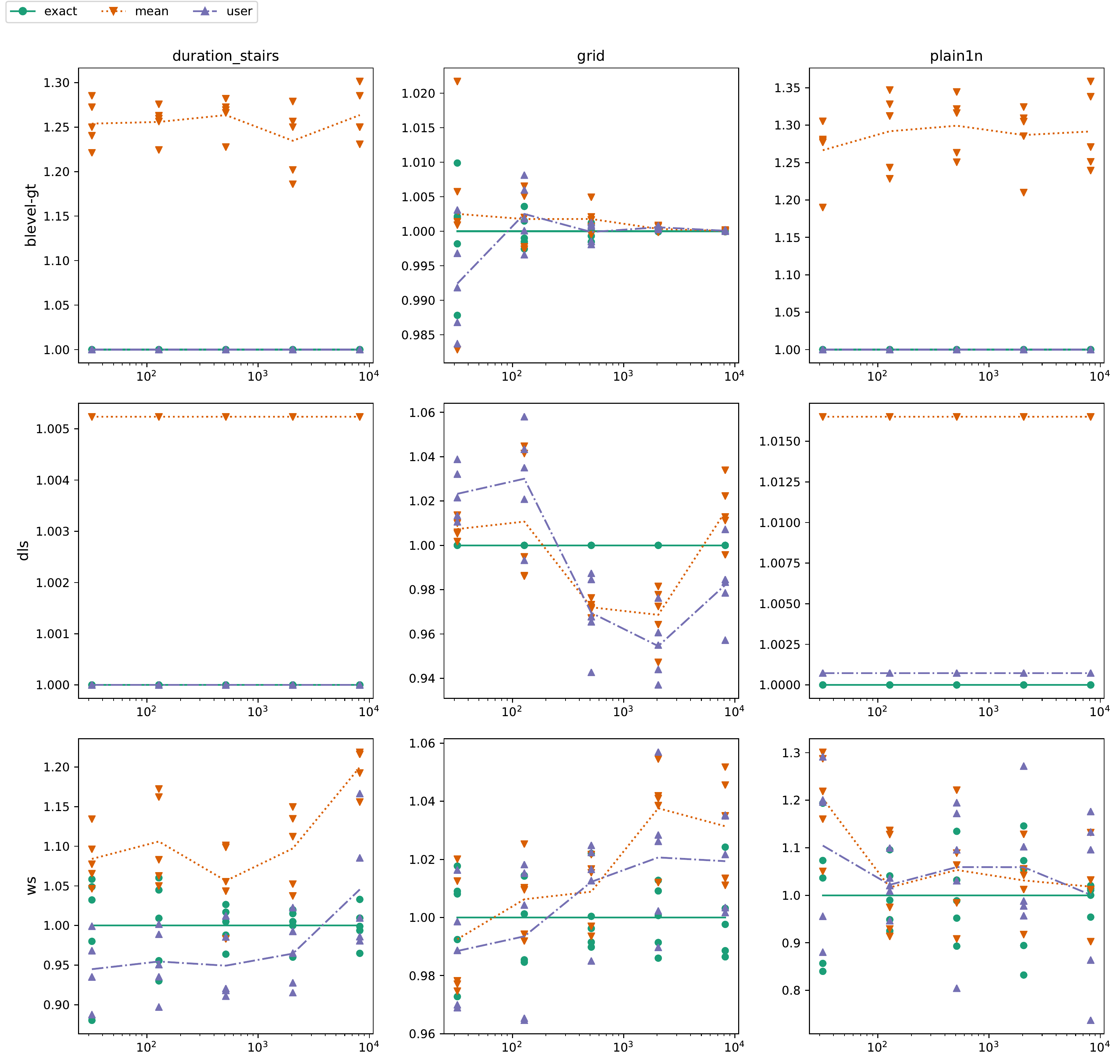}\\
{\small x axis: bandwidth [MiB/s]; y axis: makespan normalized to
	\emph{exact} average}
\end{figure}

\textbf{Imodes}\quad
Figure~\ref{f:irw-imode} compares makespans between the imodes on the IRW data
set using the 32$\times$4 cluster for selected schedulers. The results are
normalized with respect to the \emph{exact} imode.
The results show that the effect significantly depends on the particular
scheduler. The effect of imodes seems to be more relevant than the effect of
MSD, but in most cases it is still significantly smaller than the effect of the
simulated netmodel. Since the \emph{exact} imode provides the schedulers with
the most accurate and complete information that they can get, it may be
unintuitive why some schedulers actually perform better when presented with
incomplete or inaccurate data (e.g. the \emph{dls} scheduler on the
\emph{fastcrossv} graph). This is partially caused by the fact that all of the
schedulers use heuristics, they can thus produce worse results even when
presented with a more accurate input and vice versa.

Another reason is that with the \emph{max-min} netmodel, the scheduler knows
only a lower bound on the communication costs even if it knows the exact data
size in advance. It has access to the network maximum bandwidth, but does not
know the current and future network utilization, thus it only has a crude
estimation of the real transfer duration.


Figure~\ref{f:elementary-imode} shows the effect of imodes on three graphs from
the elementary set. Imode effects are mainly visible for the \emph{ws} and
\emph{blevel-gt} schedulers; for other schedulers, the effects are
significantly smaller. The task graph \emph{duration\_stairs} has tasks with
several different durations, the duration estimates produced by the \emph{mean}
imode will thus be fairly inaccurate. This is observable for the
\emph{blevel-gt} and \emph{ws} schedulers, which produce up to 25\% longer
makespans when compared to the \emph{exact} imode.

\subsection{Validation}
\label{sec:validation}
It is challenging to validate the performance of multiple task schedulers in real task
execution frameworks. Schedulers of existing task frameworks are usually very deeply integrated and
coupled to the surrounding system in order to be as performant as possible. It can thus be quite
difficult, or even infeasible, to swap the scheduler for a different one.
Task frameworks might also be fundamentally incompatible with some scheduling approaches. For
example, workstealing schedulers perform a lot of complex communications amongst workers and the
scheduler, and if the execution system does not support such communication patterns, implementing
workstealing can amount to rewriting the whole system from scratch.

We have leveraged the approach from~\citep{rsds} and used its modified version of Dask~\citep{dask}
as a validation framework. In addition to its built-in workstealing scheduler, we have
also implemented three simple scheduling algorithms into it (\emph{blevel}, \emph{tlevel} and
\emph{random}).

The absolute makespans of task graphs simulated by \estee{} and task graphs executed by some
task framework cannot be directly compared, because the framework will always introduce
runtime overheads and system noise that cannot be fully simulated. However, since one of the
goals of this paper is to compare the relative performance of various schedulers, we have
decided to compare the relative makespans normalized to a reference scheduler (\emph{blevel}) to see
if the ratios between the schedulers are similar when simulated and when executed.

To ensure that we use the same task graphs for execution and simulation,
we have executed several task graph benchmarks from~\citep{rsds} (you can find their description
in that work) in Dask and generated execution traces.
These traces were then used to reconstruct the execution times and output sizes of all tasks and
this reconstructed task graph was then simulated in \estee. We have executed the task graphs with
$24$ workers on two nodes (one with the scheduler and the second one with the workers). Each task
graph was executed and simulated three times.

\begin{figure*}
	\caption{Scheduler performance relative to \emph{blevel} in Dask and \estee}
	\label{f:estee-validation}
	\centering
\includegraphics[width=\textwidth]{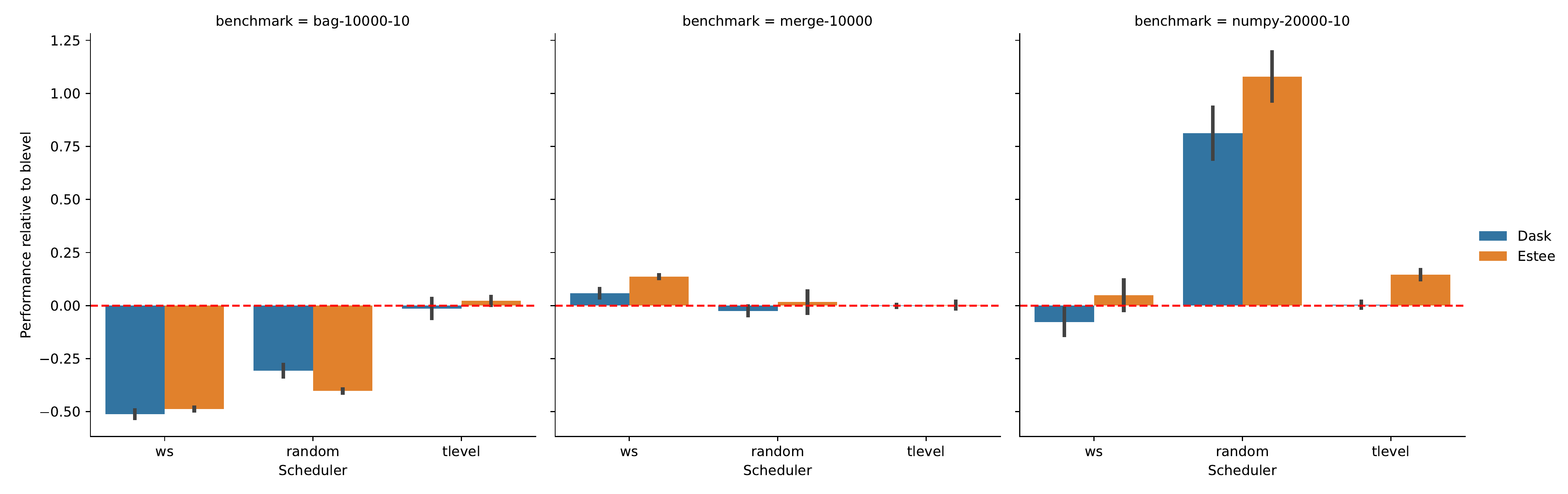}\\
{\small x axis: scheduler; y axis: performance relative to \emph{blevel}}
\end{figure*}

The performance of each scheduler was normalized to the performance of the \emph{blevel} scheduler
within the same environment. The relative ratios were centered around zero by subtracting $1$ from
them, to focus on the relative differences. For example, if a task graph took $100 s$ to execute in
Dask with the \emph{blevel} scheduler, but $110 s$ with the \emph{ws} scheduler, the ratio of
the \emph{ws} scheduler would be $0.1$. If the simulation was perfect, the two columns for each
scheduler would have the same height.

We have selected three interesting situations that can be seen in
Figure~\ref{f:estee-validation}. Full results are in the appendix in
Figure~\ref{f:estee-validation-full}.

The first chart shows a situation where changing the scheduler resulted in large changes in makespans,
and \estee{} was able to simulate these precisely. The second chart demonstrates a situation where
all schedulers produce similar makespans, therefore in this case the scheduling algorithm does not
seem to be that important. \estee{} also estimated that the differences between schedulers will
be small. In the third benchmark, \estee{} systematically overestimated the makespans
of all three schedulers with respect to the reference scheduler. While the \emph{ws} implementation
in \estee{} was partly inspired by Dask, the scheduler behaviour is quite complex and in this case
it was able to outperform the reference scheduler in a way that \estee{} wasn't able to simulate.

To summarize the average error, we took the relative makespans of individual schedulers w.r.t.\
the reference scheduler and calculated the difference between the executed and simulated
relative makespan. The geometric mean of these differences is $0.0347$, which suggests that the
differences between the execution and simulation were relatively small, usually within a few
percent.

\section{Conclusion}
\label{sec:conclusions}
We implemented a set of well known scheduling heuristics and prepared a dataset
containing workflows of different types and scales.
Based on those, we have conducted a series of fully reproducible benchmarks to
analyze the influence of network models, information modes and minimal
scheduling delays on the behavior of the implemented schedulers.

Our results show that several implementation details of both the scheduling algorithms and
the simulation environment must be clearly described and specified, otherwise the results
might not be reproducible. We have shown that the complexity of the used network model
may significantly affect the simulated workflow execution makespan.
To our surprise, the effect of information modes has been relatively low for
most of the benchmarked cases.
It seems that for the benchmarked scheduling algorithms, it is relatively
sufficient to know only rough estimates of task durations and data object sizes.

Lastly, we showed that various MSD values have a limited impact on the
resulting makespan, but increasing the scheduling delay may in some cases
improve the produced schedules.


Our results confirmed that it is important to consider the network behavior
when applying scheduling heuristics in real-world applications and that it
requires caution to refer to results that use simplified network models.
We also encourage authors of scheduling algorithms to describe the worker selection
strategy, possible delay between scheduler invocations, network model and other
implementation details in utmost detail, to make scheduler benchmarks reproducible.

\estee{}, workflow datasets and scheduler implementations are open sourced, to
make the results reproducible and extendable by the community. We believe that
our results provide a comprehensive overview and comparison of workflow
schedulers in various simulated conditions and that \estee{} has further
potential to simplify the development and benchmarking of novel schedulers.

\section*{Acknowledgements}
This work was supported by the ACROSS project. This project has received funding from the European
High-Performance Computing Joint Undertaking (JU) under grant agreement No 955648. The JU receives
support from the European Union’s Horizon 2020 research and innovation programme and Italy, France,
the Czech Republic, the United Kingdom, Greece, the Netherlands, Germany, Norway. This project has
received funding from the Ministry of Education, Youth and Sports of the Czech Republic (ID: MC2104).

This work was supported by the Ministry of Education, Youth and Sports of the Czech Republic through
the e-INFRA CZ (ID:90140).

\backmatter

\bibliography{references}

\appendix
\section{W-scheduler}
\label{sec:w-sched}
This section contains description of the \emph{w-scheduler} -- worker inner
scheduler.
The (global) scheduler does not directly communicate with w-schedulers except
by assigning tasks to workers. The assignment of task $t$ may also contain two
additional values: priority $p_t$ and blocking $b_t$, such that $b_t \leq p_t$.
These values set priorities for downloading and task execution when more
possible options are enabled at once for the w-scheduler.

Worker $w$ starts to download an input $o \in \outputs$ for a task $t \in
\tasks$ if $t$ was assigned on $w$ by the scheduler and $o$ is not already on
$w$. The download is started as soon as the task producing $o$ is finished and
there is a free download slot.

When more objects can be downloaded at once but there are not enough download
slots, the downloads are prioritized based on the priority of tasks that need
the object. When a task is not ready then its $p$ is used, otherwise $p$ is
boosted by a constant.
When more tasks need the same object, then the maximum priority is taken.
Downloading is uninterruptible, once an object has started downloading, it is
finished without interruption even when a download with a higher priority is
enabled and the maximum number of concurrent downloads per worker is reached.

Download slots serve to limit simultaneous downloads. For the \emph{max-min}
network model, the worker is allowed to download at most four inputs at once,
but at most two from the same worker. These particular numbers were observed as
a reasonable compromise between parallel downloads and using bandwidth for
higher priority tasks. For the \emph{simple} model, we allow the worker to run
arbitrarily many simultaneous downloads to make the model behave in a way that
is similar to previous studies.

When a new task becomes enabled on a worker $w$ or an execution of a task is
finished,
worker runs the following algorithm to decides if another task can be executed.
We denote $f$ as the a of free CPU cores (i.e. the total number of worker's
cores minus the sum of core requirements of currently running tasks),
$E$ as a set of tasks that are enabled and non-running and $X$ as a set of
tasks from $E$ which require more than $f$ CPU cores.
The worker picks a task $t$ from $E \setminus X$ with maximal priority such
that $\forall t' \in X: p_t \leq b_{t'}$. If such $t$ exists, then $t$ is
started. This process is repeated until we cannot start another task this way.

\newpage
\section{Benchmark results}
\label{sec:benchmark-results}
\begin{figure*}[b]
	\caption{Complete scheduler comparison on IRW set}
	\label{f:irw-s-t-full}
	\centering
\includegraphics[width=0.9\textwidth]{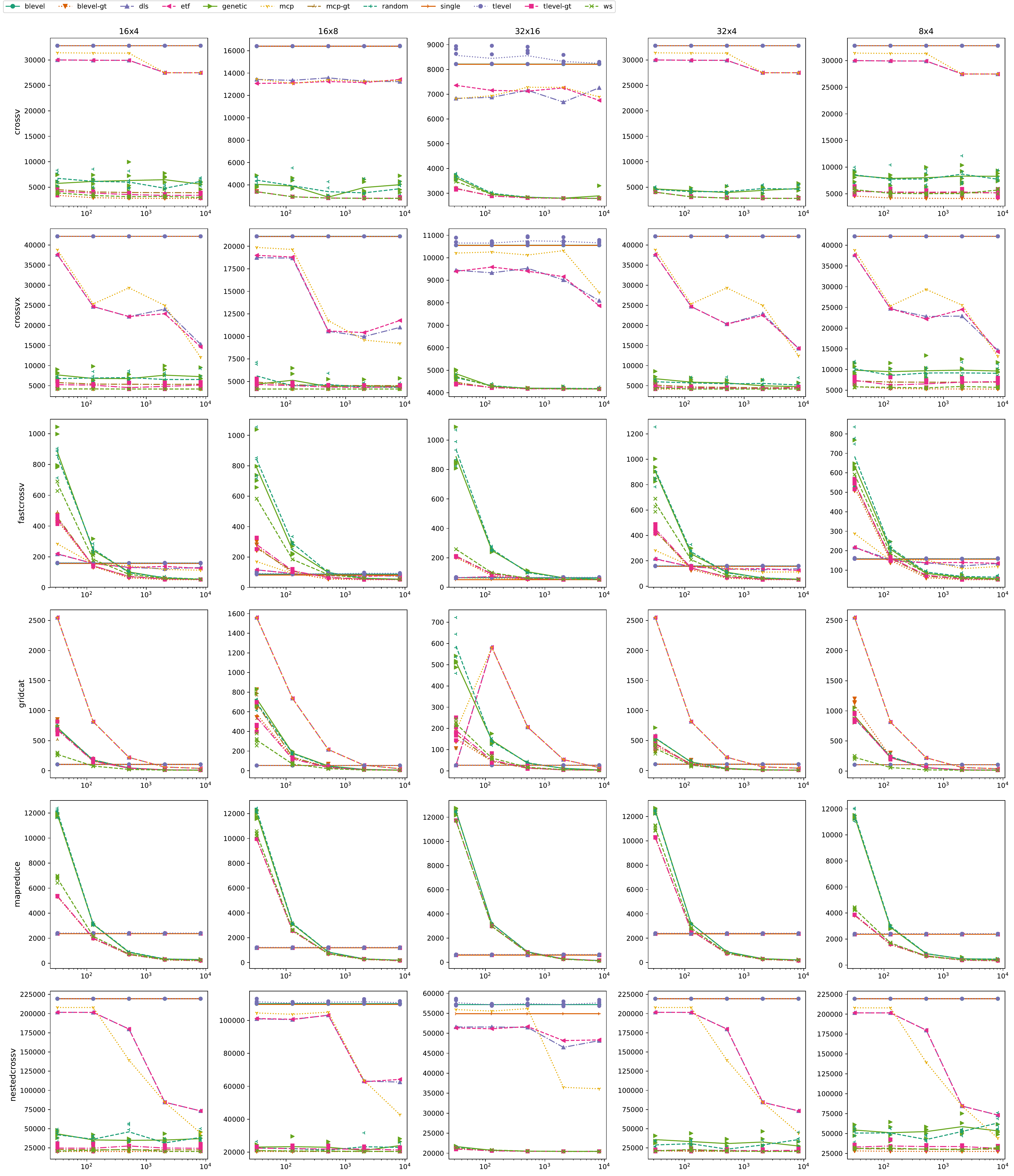}\\
{\small x axis: bandwidth [MiB/s]; y axis: makespan [s]}
\end{figure*}

\begin{figure*}
	\caption{Comparison of ``maxmin'' and ``simple'' netmodel on Pegasus set;
cluster 32x4}
	\label{f:pg-netmodel}
	\centering
	\includegraphics[width=\textwidth]{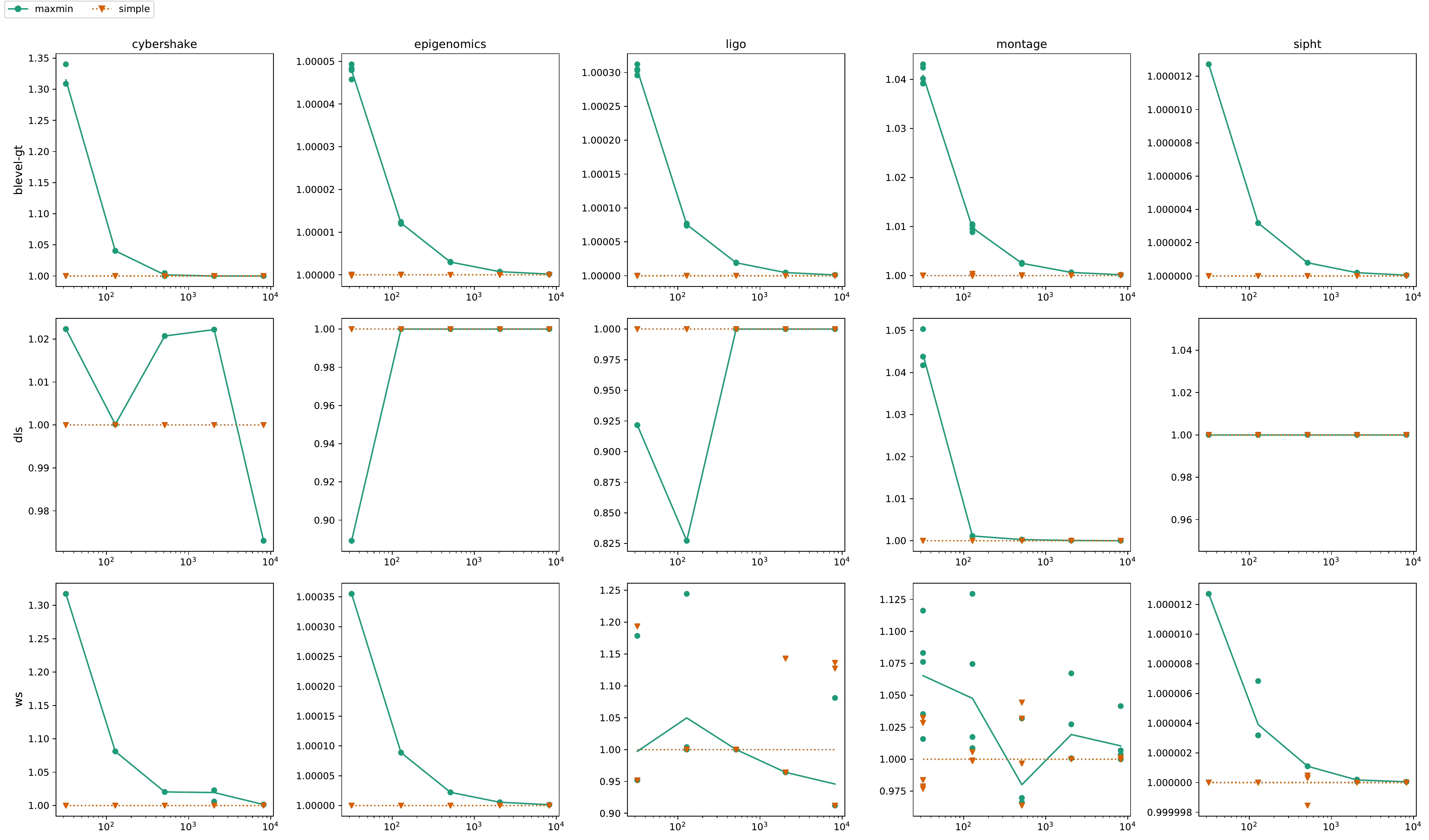}
	\\ {\small x axis: bandwidth [MiB/s]; y axis: execution makespan normalized
to average of makespan of ``simple'' model}
\end{figure*}

\begin{figure*}
	\caption{Scheduler performance relative to \emph{blevel} in Dask and \estee}
	\label{f:estee-validation-full}
	\centering
\includegraphics[width=\textwidth]{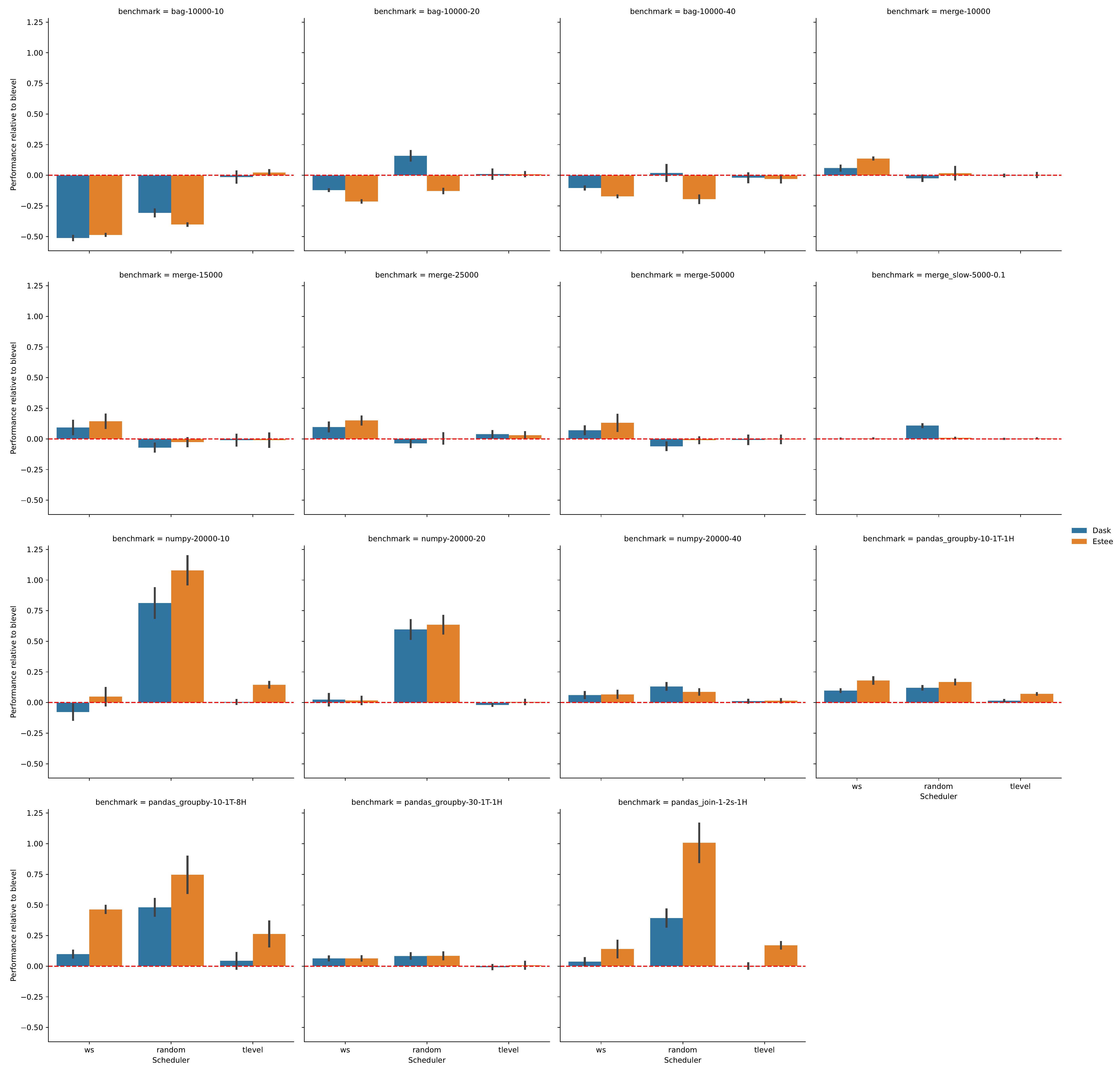}\\
{\small x axis: scheduler; y axis: performance relative to \emph{blevel}}
\end{figure*}

\end{document}